\RequirePackage{ifpdf}
\ifpdf 
	\documentclass[twocolumn, showpacs, showkeys, footinbib, amsmath, amssymb, prl, letterpaper, pdftex]{revtex4}
\else
	\documentclass[twocolumn, showpacs, showkeys, footinbib, amsmath, amssymb, prl, letterpaper, dvips]{revtex4}
\fi

\usepackage[nodvipsnames]{color}
\definecolor{green}		{rgb}{0., 	0.5, 	0.}
\definecolor{violet}	{rgb}{0.5, 	0., 	0.5}
\definecolor{gray}		{rgb}{0.5, 	0.5, 	0.5}
\definecolor{blue}		{rgb}{0., 	0., 	0.5}
\definecolor{charcoal}	{rgb}{0.2, 	0.2,	0.2}
\definecolor{brown}		{rgb}{0.6, 	0.2, 	0.}
\definecolor{teal}		{rgb}{0.,	0.5, 	0.5}

\usepackage{dcolumn}
\newcolumntype{.}{D{.}{.}{-1}}
\newcolumntype{1}{D{.}{.}{1}}
\newcolumntype{2}{D{.}{.}{2}}
\newcommand{\rb}[1]{\raisebox{-1.5ex}[0pt][0pt]{#1}}
\newcommand{\mc}[1]{\multicolumn{1}{c}{#1}}
\newcommand{\mcTwo}[1]{\multicolumn{2}{c}{#1}}
\newcommand{\mcThree}[1]{\multicolumn{3}{c}{#1}}
\newcommand{\mcFour}[1]{\multicolumn{4}{c}{#1}}
\newcommand{\mcFive}[1]{\multicolumn{5}{c}{#1}}

\newcommand{\cHat}{\ensuremath{\hat{c}}}
\newcommand{\mOne}{\ensuremath{_\text{max1}}}
\newcommand{\mTwo}{\ensuremath{_\text{max2}}}
\newcommand{\agglo}{\ensuremath{_\text{a}}}
\newcommand{\x}{\ensuremath{_\text{x}}}
\newcommand{\y}{\ensuremath{_\text{y}}}

\ifpdf
  \usepackage[pdftex]{graphicx}
  \usepackage{subfigure}
  \graphicspath{{pdf-figures/}}
  \usepackage[%
    pdftex,%
    pdftitle = {{Grain-boundary grooving and agglomeration in alloy thin films}},%
	pdfauthor = {{Mathieu Bouville, Dongzhi Chi, and David J. Srolovitz}},%
	pdfsubject = {{As the groove grows, the slow species becomes concentrated near the groove tip, which limits further grooving. At early times the dominant diffusion path is along the boundary, while at late times it is parallel to the substrate. This change in path strongly affects the time-dependence of grain boundary grooving and increases the time to agglomeration.}},%
	pdfkeywords = {{phase-field method, agglomeration}},%
	pdfpagemode = UseOutlines,%
  ]{hyperref}
\else
  \usepackage[dvips]{graphicx}
  \usepackage{subfigure}
  \usepackage[%
    dvips,%
    pdftitle = {{Grain-boundary grooving and agglomeration of alloy thin films with a slow-diffusing species}},%
	pdfauthor = {{Mathieu Bouville, Dongzhi Chi, and David J. Srolovitz}},%
	pdfsubject = {{As the groove grows, the slow species becomes concentrated near the groove tip, which limits further grooving. At early times the dominant diffusion path is along the boundary, while at late times it is parallel to the substrate. This change in path strongly affects the time-dependence of grain boundary grooving and increases the time to agglomeration.}},%
	pdfkeywords = {{phase-field method, agglomeration}},%
	pdfpagemode = UseOutlines,%
  ]{hyperref}
\fi

\begin{document}

\title[Grooving and agglomeration of alloy thin films]{Grain-boundary grooving and agglomeration of alloy thin films\\with a slow-diffusing species}

\author{Mathieu Bouville}
	\email{m-bouville@imre.a-star.edu.sg}
	\affiliation{Institute of Materials Research and Engineering, Singapore 117602}

\author{Dongzhi Chi}%
	\affiliation{Institute of Materials Research and Engineering, Singapore 117602}

\author{David J. Srolovitz}%
	\affiliation{Department of Physics, Yeshiva University, New York, NY 10033, USA}

\date{\today}

\begin{abstract}
We present a general phase-field model for grain-boundary grooving and agglomeration of poly\-crystalline alloy thin films. In particular, we study the effects of slow-diffusing species on grooving rate. As the groove grows, the slow species becomes concentrated near the groove tip so that further grooving is limited by the rate at which it diffuses away from the tip. At early times the dominant diffusion path is along the boundary, while at late times it is parallel to the substrate. This change in path strongly affects the time-dependence of grain boundary grooving and increases the time to agglomeration. The present model provides a tool for agglomeration-resistant thin film alloy design.
\end{abstract}


\maketitle

\begin{figure}
\centering
\setlength{\unitlength}{1cm}
\begin{picture}(8.5,2.2)(.1,0)
\shortstack[c]{
\subfigure{
	\label{basic-grooving(a)}
	\includegraphics[width=4.2cm]{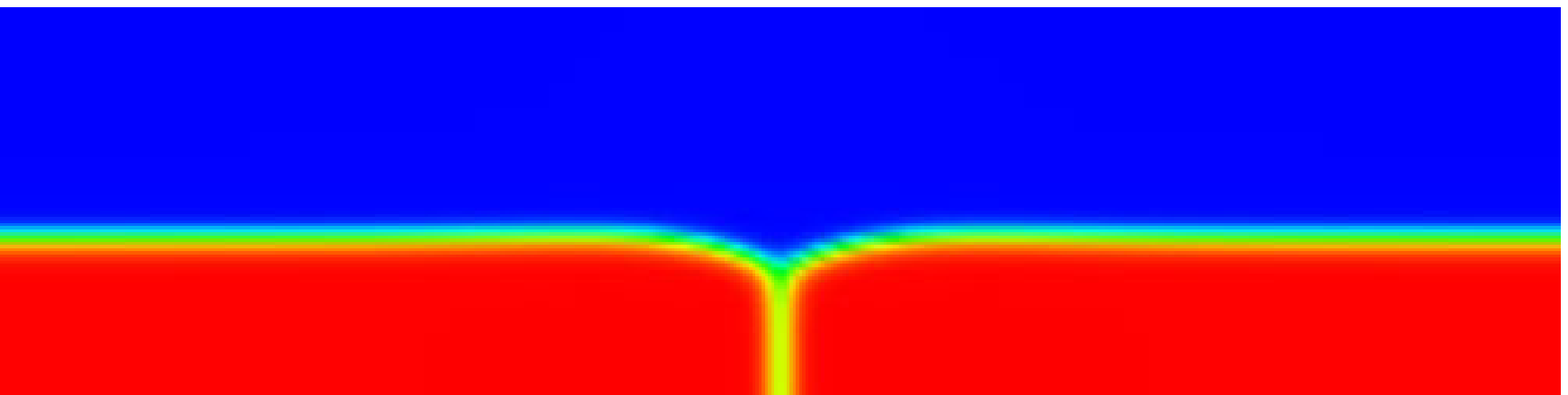}
	\put(-0.55,0.65){\color{white}\bf{(a)}}
	\put(-3.8,0.6){\color{white}Si}
	\put(-3.95,0.1){\color{white}NiSi}
}\subfigure{
	\label{basic-grooving(b)}
	\includegraphics[width=4.2cm]{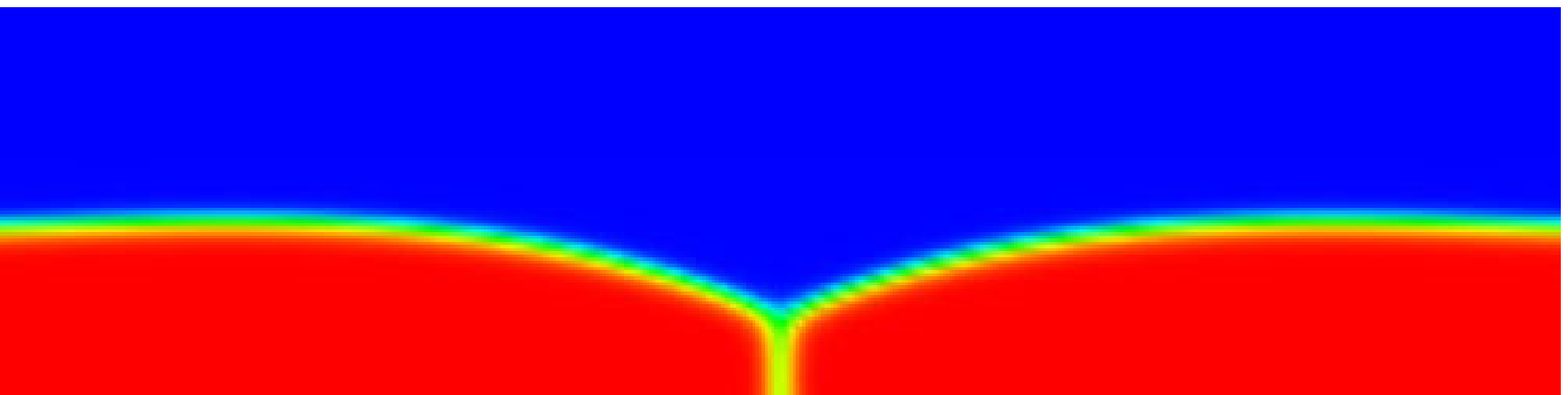}
	\put(-0.55,0.65){\color{white}\bf{(b)}}
}\vspace{-3.5mm}\\
\subfigure{
	\label{basic-grooving(c)}
	\includegraphics[width=4.2cm]{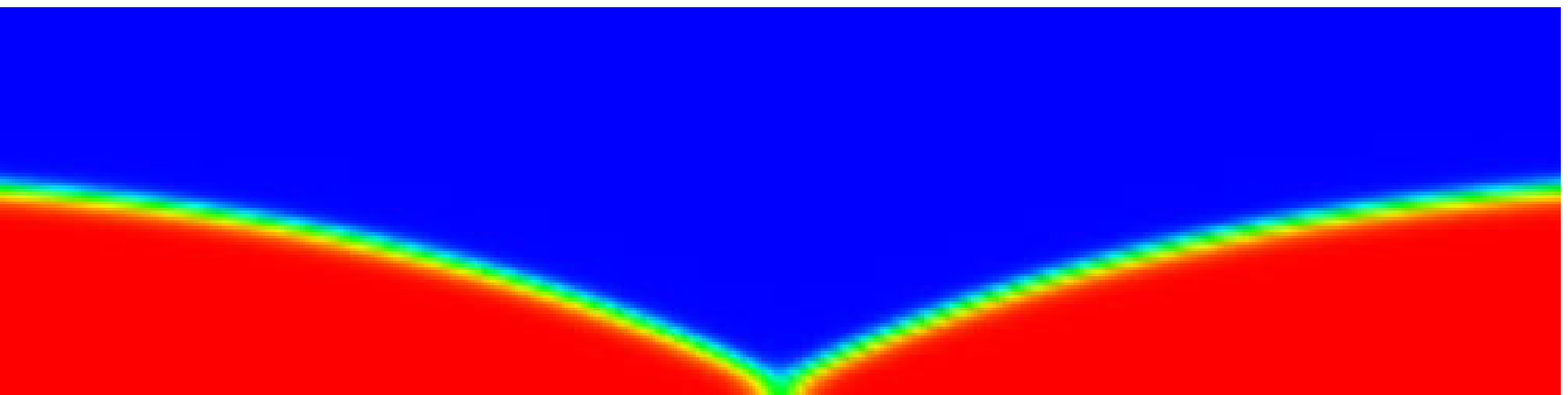}
	\put(-0.55,0.65){\color{white}\bf{(c)}}
}\subfigure{
	\label{basic-grooving(d)}
	\includegraphics[width=4.2cm]{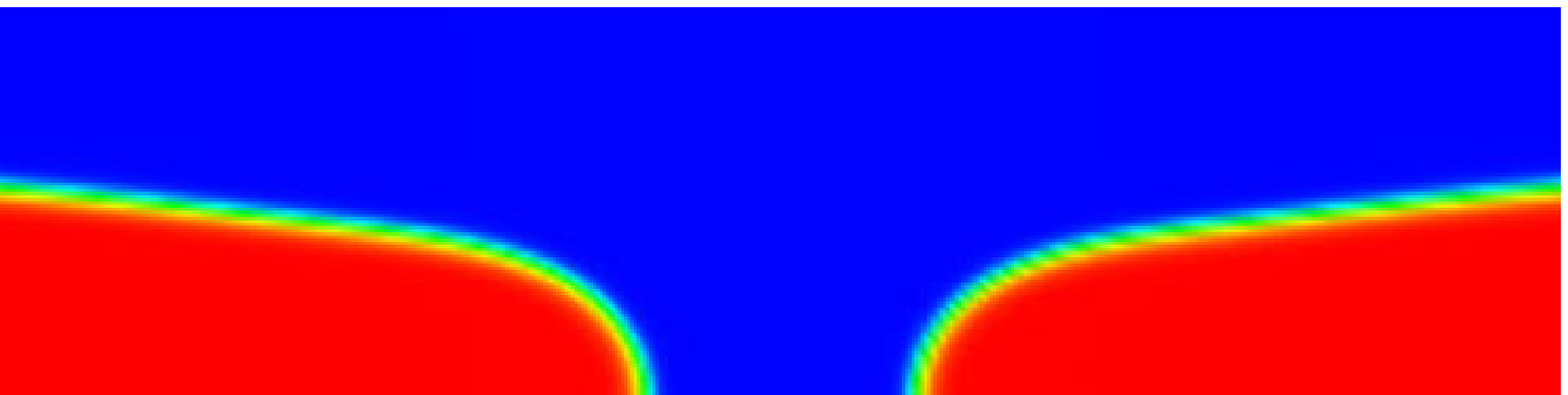}
	\put(-0.55,0.65){\color{white}\bf{(d)}}
}}
\end{picture}
\caption{\label{basic-grooving} (color online) The time evolution of the film morphology for $h=10$~nm and $r=1$. (a) $t=0.15$~s, (b) $t=7.5$~s, (c) $t=63$~s, and (d) $t=64.5$~s.}
\end{figure}

A common failure mode in polycrystalline thin films is grain-boundary grooving through the thickness of the film. The grain-boundary shrinks while the surface or substrate/film interface extends in order to reduce the total interface energy (see Fig.~\ref{basic-grooving}). At equilibrium, the surface and interface shapes must correspond to a constant chemical potential. For an isotropic interface energy, this means constant curvature, i.e.\ a circular arc in two dimensions (2D). The interface thus evolves towards a constant curvature shape with a deepening grain-boundary groove (subject to fixed groove angle~\cite{Bailey-PPSB-50, Mullins-JAP-57} and grain volume constraints).
This can bring the surface in contact with the substrate, leading to film agglomeration~\cite{Srol-JOM-95}, as shown in Fig.~\ref{basic-grooving(d)}. Many polycrystalline metal silicide films fail in this manner during post-reaction annealing~\cite{Nolan-JAP-92, Pramanick-APL-93}. 

Grain-boundary grooving has received a great deal of attention over the past half-century~\cite{Bailey-PPSB-50, Mullins-JAP-57, srol-JAP-86-I, miller-JMR-90, Genin-acta-93, thouless-acta-93, Ramasubramaniam-acta_mat-05, Xin-acta_mat-03, Zhang-acta_mat-04}. 
However, the extant models for grain-boundary grooving and thin film agglomeration are too idealized to be useful to predict agglomeration in most technologically interesting materials, such as multicomponent alloy films. In particular, few models consider the complexity of real grain-boundary grooving --- simultaneous bulk, grain-boundary, interface, and surface diffusion;  phase transformations; etc. In this Letter, we use the phase-field method to study grain-boundary grooving and the agglomeration of a polycrystalline alloy thin film where each alloy component has a unique diffusivity and a phase transformation occurs.   For concreteness, we choose a polycrystalline NiSi film on a silicon substrate as our prototype system. We examine the effect of introducing a slow-diffusing species (Pt) into the NiSi thin film.

In our phase-field calculations, the concentrations of silicon, nickel, and platinum are $c_\text{Si}$, $c_\text{Ni}$, and $c_\text{Pt}$ respectively; all three fields are conserved. $\phi$ indicates the local fraction of the silicon phase present and $\psi_i$ the local fraction of grain $i$ of the compound silicide phase; these are not conserved. 
The free energy of the system is given by~\cite{Bouville-MSMSE-06}
\begin{eqnarray}
	&G\!& = \int_V \big[
	W\,(\psi_1)^2 (1-\psi_1)^2 +W\,(\psi_2)^2 (1-\psi_2)^2\nonumber\\
	&&{}+W \phi^2 (1-\phi)^2\! 
	+X_\text{I} (\psi_1+\psi_2)^2\phi^2\! + X_\text{GB} (\psi_1)^2 (\psi_2)^2\nonumber\\
	&&{}+A\,(c_\text{Si} - C_\text{IV})^2 + A\,(c_\text{Ni} + c_\text{Pt} - C_\text{M})^2\nonumber\\
	&&{}+\frac{k_\text{B}\,T}{\Omega} \left(\frac{c_\text{Ni}}{C_\text{M}} \,\ln \frac{c_\text{Ni}}{C_\text{M}}
	+\frac{c_\text{Pt}}{C_\text{M}} \,\ln \frac{c_\text{Pt}}{C_\text{M}}\right)\nonumber\\
	&&{}+ \lambda_\text{Si} \|\mathbf{\nabla} c_\text{Si}\|^2
	+ \lambda_\text{Ni} \|\mathbf{\nabla} c_\text{Ni}\|^2 
	+ \lambda_\text{Pt} \|\mathbf{\nabla} c_\text{Pt}\|^2\nonumber\\
	&&{}+\kappa (\|\mathbf{\nabla} \psi_1\|^2 + \|\mathbf{\nabla} \psi_2\|^2 + \|\mathbf{\nabla} \phi\|^2)
\big]\, dV.
\label{G_equation}
\end{eqnarray}
\noindent In Eq.~(\ref{G_equation}), the integral is over the entire volume of the system, $V$. The first three terms are double-well potentials for $\psi_i$ and $\phi$, which ensure that in the bulk $\psi_i$ and $\phi$ are either zero or one. The fourth and fifth terms ensure that at any point in the system, only a single phase is present. The third line of Eq.~(\ref{G_equation}) fixes the stoichiometry (in the substrate $C_\text{M}=0$ and $C_\text{IV}=1$, whereas in the silicide film $C_\text{M}=C_\text{IV}=1/2$), while the fourth line describes the configurational entropy associated with the metal sublattice of the silicide. The gradient terms in the last two lines enforce smooth interfaces.  The parameters $W$\!, $X_I$, $X_{GB}$, $A$, $\kappa$, and $\lambda_i$ set the overall thermodynamics of the film ($\Omega$ is the average atomic volume). 

The temporal evolution of the compositions and phases is related to the free energy $G$, respectively through the Cahn--Hillard equation~\cite{Cahn-acta_met-61}
\begin{equation}
	\frac{\partial\, c(\mathbf{r}, t)}{\partial\, t} = \mathbf{\nabla} M\, \mathbf{\nabla} \frac{\delta\, G}{\delta\, c(\mathbf{r}, t)}
\label{Cahn-Hillard}
\end{equation}
\noindent and the Allen--Cahn equation~\cite{Allen-Jphys-77}
\begin{equation}
	\frac{\partial\, \psi(\mathbf{r}, t)}{\partial\, t} = -L \frac{\delta\, G}{\delta\, \psi(\mathbf{r}, t)}.
\label{Allen-Cahn}
\end{equation}
\noindent Here $M$ and $L$ are atom and interface mobilities, respectively, and $\delta$ indicates a functional derivative.

\begin{figure}
\centering
	\includegraphics[width=8.5cm]{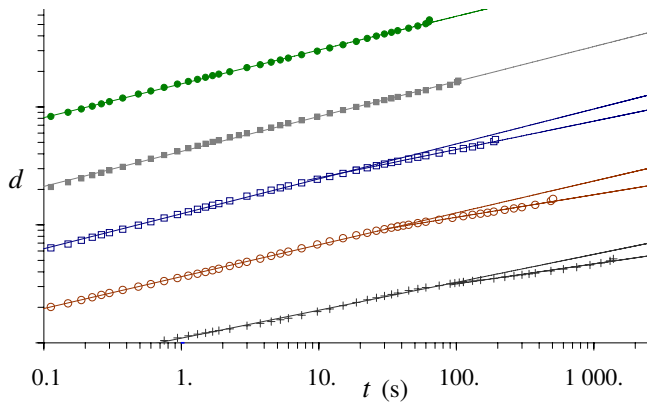}
\caption{\label{t-d-offset} The depth of the groove for $h=10$~nm as a function of time (log--log scale).
{\color{green}$\bullet$}:~$r		= 1$, 
{\color{gray}$\blacksquare$}:~$r	= 3\times10^{-2}$,
{\color{blue}$\Box$}:~$r			= 10^{-2}$,
{\color{brown}$\circ$}:~$r			= 3\times10^{-3}$, and 
{\color{charcoal}$+$}:~$r			= 10^{-3}$. 
The curves are offset by a half decade, for clarity. The lines are power law fits to Eq.~(\ref{eq-depth}) with parameters as shown in Table~\ref{table-alpha}.}
\end{figure}

The two-dimensional system consists of two silicide grains on a silicon substrate, discretized on a  $512 \times 96$ mesh.  For simplicity, we consider the symmetric case where the phase above the film is also silicon (rather than vacuum).  This allows us to use periodic boundaries along both directions, as well as fewer parameters in the free energy (e.g.\ the surface and interface are identical).
The evolution equations, Eqs.~(\ref{Cahn-Hillard}) and~(\ref{Allen-Cahn}), are solved in reciprocal space using a semi-implicit method, as described in Ref.~\onlinecite{Chen-98}. 
We use the following values for the parameters: $W=5$, $X_\text{I}=30$, $X_\text{GB}=18$, $A = 8$, $k_\text{B} T/\Omega= 0.3$, $\lambda_\text{Si} = 20$, $\lambda_\text{Ni} = 30$, $\lambda_\text{Pt} = 100$, $\kappa=8$, $L=4$, and $M=1$. We choose the length and time scales to represent a typical case of a 10~nm-thick NiSi film on Si, which is experimentally observed to agglomerate in around a minute at $T=600$~\mbox{$^\circ$C}. Film thicknesses are thus given in nanometers for convenience, not based on the interface thickness. One should note that our simulations are relevant to grain-boundary grooving and agglomeration of alloy thin films in general and that ratios of times and distances are more important than absolute values.

This Letter focuses on how the introduction of a slow-diffusing species into the film affects grain-boundary grooving and agglomeration. We perform a series of simulations at different ratios of the Pt to Ni mobilities, $r=M_\text{Pt} / M_\text{Ni} \le 1$ (we keep the Ni mobility fixed), spanning three orders of magnitude.  We note that the silicide is a compound in which the Pt substitutes for Ni on the metal sublattice. In the present simulations, the average concentration of Pt is 7\%.

\begin{table}
\begin{tabular}{l221l221l1l1r}
\hline
\mc{\rb{$r$}}		&\mcTwo{early}	&\mcTwo{}	&\mcTwo{late}			&\mcFour{agglomeration}\\
\mc{}			&\mc{$d_0$}&\mc{$\alpha$}	&\mcTwo{$t^*$} &\mc{$d_0$}&\mc{$\alpha$}&\mcTwo{$t\agglo$}& \mcTwo{$\cHat\agglo$}\\
\hline
$1\!\times\!10^0$\rule[0ex]{0ex}{2.6ex}& 2.37 & 0.30	& \mcTwo{---}& \mc{---} & \mc{---}	& 1.0	&min		& 8.0&\%	\\
$1\!\times\!10^{-1}$	& 2.22	& 0.30	& \mcTwo{---}& \mc{---} & \mc{---}	& 1.2	&min	& 8.0&\%	\\
$3\!\times\!10^{-2}$	& 2.13	& 0.29	& \mcTwo{---}& \mc{---} & \mc{---}	& 1.7	&min	& 8.2&\%	\\
$1\!\times\!10^{-2}$	& 1.96	& 0.30	& 14.4	&s	 & 2.28		& 0.24		& 3.2	&min	& 8.5&\%	\\
$3\!\times\!10^{-3}$	& 1.81	& 0.27	& 35	&s	 & 2.47		& 0.18		& 8.4	&min	& 8.7&\%	\\
$1\!\times\!10^{-3}$	& 1.74	& 0.24	&  1.2	&min & 2.36		& 0.17		& 23	&min	& 8.7&\%	\\
\hline
\end{tabular}
\caption{\label{table-alpha} The grooving parameters $d_0$ (in nm~s$^{-\alpha}$) and $\alpha$ from Eq.~(\ref{eq-depth}) ---as determined by fits in Fig.~\ref{t-d-offset}---, $t\agglo$, the time at which the film agglomerates, and $\cHat\agglo$, the value of $\cHat$ at agglomeration 
for $h=10$~nm. For small values of $r$, we quote early and late time values of $d_0$ and $\alpha$ corresponding to the two regimes shown in Fig.~\ref{t-d-offset}. $t^*$ is the time at which the early and late time fits intersect in Fig.~\ref{t-d-offset}.}
\end{table}

Figure~\ref{t-d-offset} shows the groove depth~$d$ as a function of time~$t$ for several values of $r$. The time-dependence of the groove depth is of the form
\begin{equation}
	d = d_0 \,t^\alpha.
	\label{eq-depth}
\end{equation}
\noindent \Citet{Mullins-JAP-57} found that $\alpha=1/4$ if mass transfer occurs by surface diffusion and $\alpha=1/2$ if by evaporation and condensation. At early times the groove depth is well described by Eq.~(\ref{eq-depth}), with $\alpha$ between 0.24 and 0.30 (see Table~\ref{table-alpha}), in reasonable agreement with the theoretical exponent for the surface diffusion case. The prefactor $d_0$ increases with $r$; as expected, the presence of a slow-diffusing species in the film retards grain-boundary grooving.
However, Fig.~\ref{t-d-offset} shows that at late times the apparent value of $\alpha$ decreases to a lower, constant value. This cross-over time $t^*$ is larger for smaller $r$. A change of the value of $\alpha$ has also been found in Refs.~\onlinecite{Genin-acta-93} and~\onlinecite{thouless-acta-93}.

Clearly, decreasing the mobility of one of the species in the film  (Pt here) has a major effect on the time required for agglomeration to occur. Table~\ref{table-alpha} shows that decreasing $r$ from $1$ to $r = 10^{-2}$ triples the time required for film agglomeration, while decreasing $r$ to $10^{-3}$ leads to a twenty-fold increase. Although introduction of a slow-diffusing species was expected to slow grain-boundary grooving and delay agglomeration, the complex interplay between morphology and composition in multicomponent systems adds unforeseen richness.   An alloy with a slow-diffusing species does not behave like a monocomponent system with a lower diffusivity, as often assumed. 

\begin{figure}
\centering
\setlength{\unitlength}{1cm}
\begin{picture}(8.5, 6)(.1,0)
\shortstack[l]{
\subfigure{
	\label{slow_diff0}
	\includegraphics[width=7.7cm]{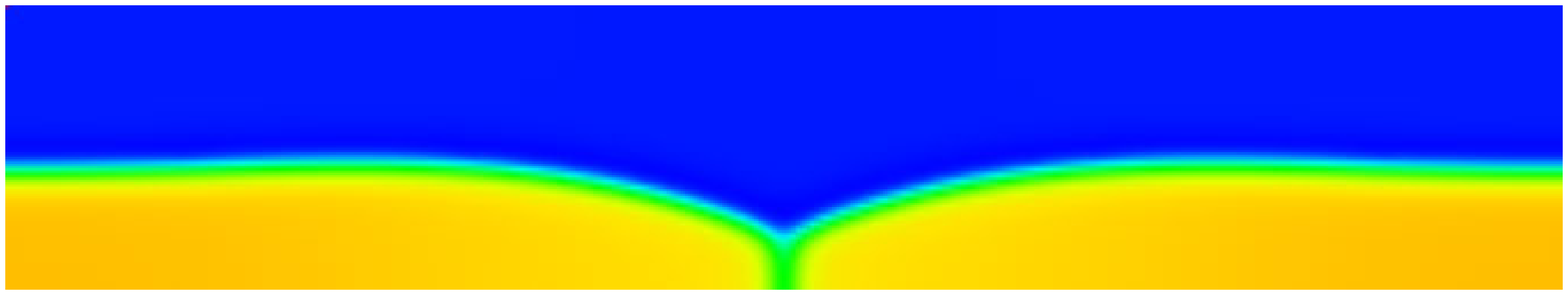}
	\put(-0.6,1){\color{white}\bf{(a)}}
}\vspace{-3mm}\\
\subfigure{
	\label{slow_diff1}
	\includegraphics[width=7.7cm]{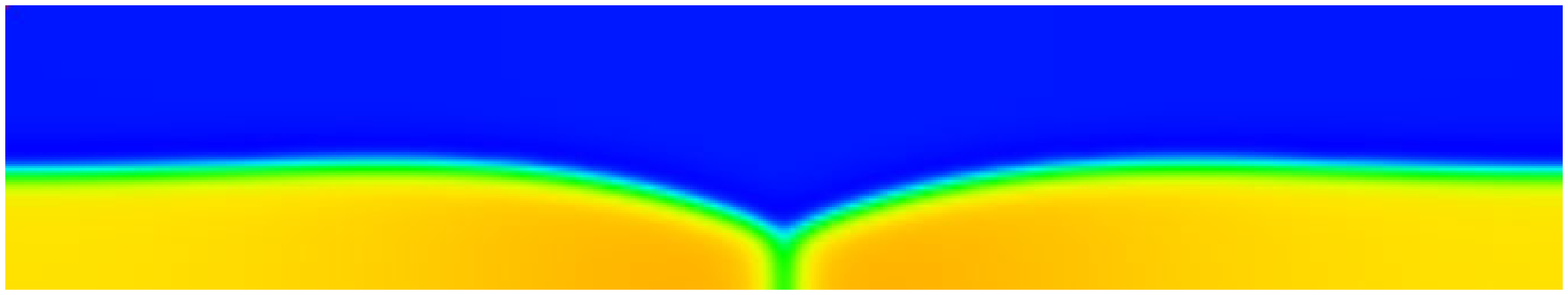}
	\put(-0.6,1){\color{white}\bf{(b)}}
}\vspace{-3mm}\\
\subfigure{
	\label{slow_diff2}
	\includegraphics[width=7.7cm]{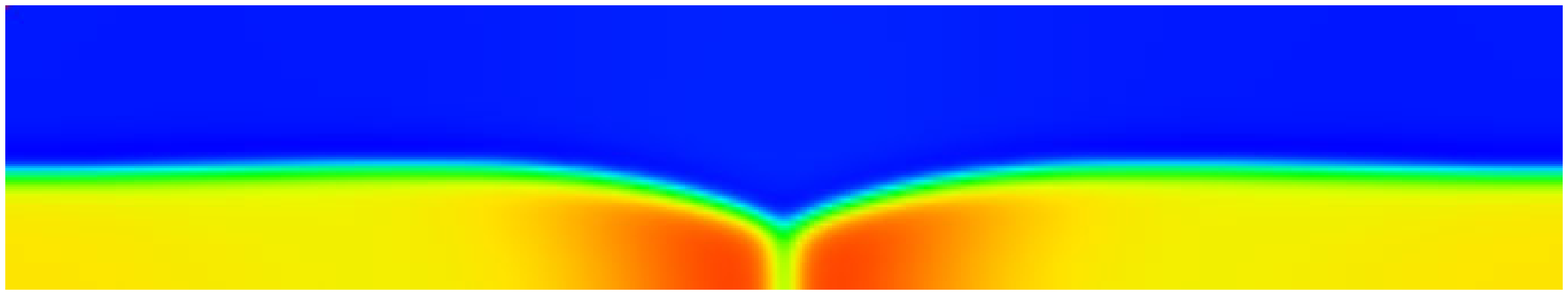}
	\put(-0.6,1){\color{white}\bf{(c)}}
}\vspace{-3mm}\\
\subfigure{
	\label{slow_diff3}
	\includegraphics[width=7.7cm]{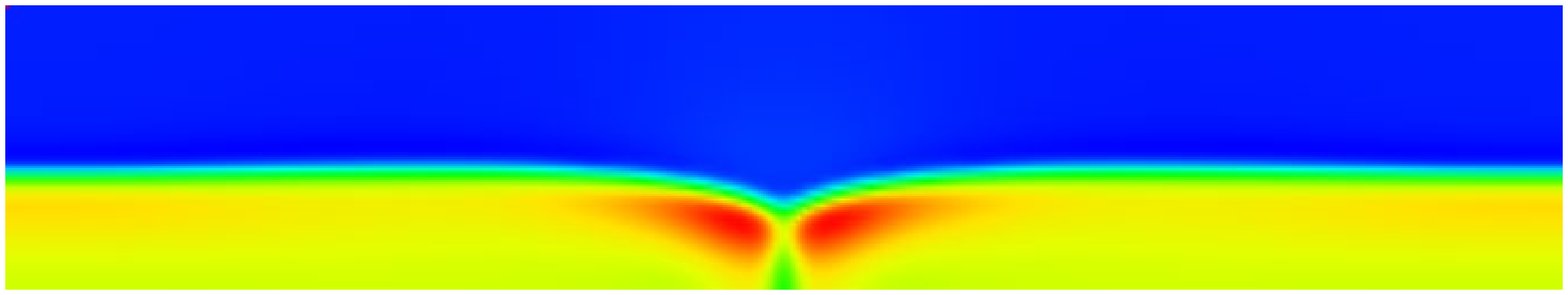}
	\put(-0.6,1){\color{white}\bf{(d)}}
}}\;\includegraphics[height=3.cm]{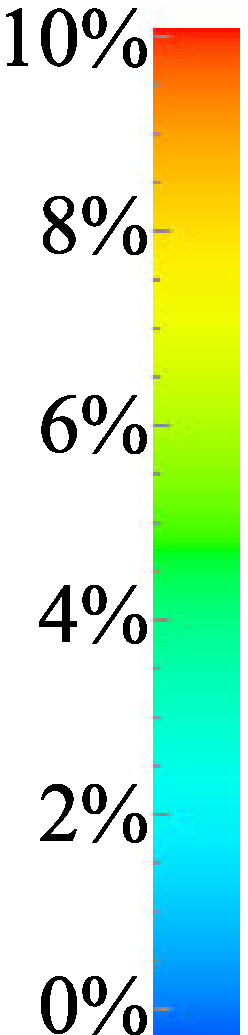}
\end{picture}
\caption{\label{slow_diff} (color online) The Pt composition field at $t = 9$~s for $h=10$~nm and several values of $r = M_\text{Pt} / M_\text{Ni}$: (a) $r = 1$, (b) $r = 10^{-1}$, (c) $r = 10^{-2}$, and (d) $r = 10^{-3}$.}
\end{figure}

Figure~\ref{slow_diff} shows that for small $r$ there is a notable excess of platinum near the grain-boundary/interface intersection;  the Pt content close to the groove is roughly 40\% larger than the average Pt concentration in the film. Indeed the groove can grow only as fast as the metal atoms can diffuse away from the deepening groove.  When Pt diffuses much more slowly than Ni, the grooving rate is limited by the time required for the advancing groove to `push the Pt out of the way.'  On the other hand, if $r=1$  or $r=10^{-1}$ (see Fig.~\ref{slow_diff}), the Pt composition is nearly homogeneous throughout the film. In this case, the alloy behaves similarly to a monocomponent system and the time evolution of the groove depth follows Eq.~(\ref{eq-depth}). This is consistent with the observations made above for the effect of Pt on the rate of grooving (Fig.~\ref{t-d-offset} and Table~\ref{table-alpha}) and with experimental observations that show that the introduction of Pt into nickel silicide greatly slows grooving and agglomeration~\cite{Lee-microelec-02}.

\begin{figure}
\centering
	\includegraphics[width=8.5cm]{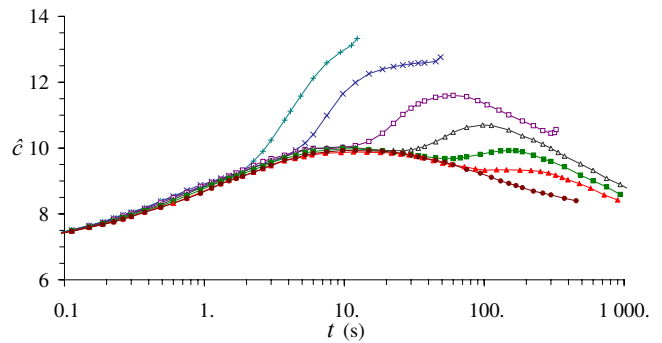}
\caption{\label{t-cMax-slow3} $\cHat$ as a function of time for $r=10^{-3}$.
{\color{teal}$+$}:~$h				=5$~nm,
{\color{blue}$\times$}:~$h			=6$~nm,
{\color{violet}$\Box$}:~$h			=8$~nm,
{\color{charcoal}$\triangle$}:~$h	=10$~nm, 
{\color{green}$\blacksquare$}:~$h	=12$~nm, 
{\color{red}$\blacktriangle$}:~$h	=14$~nm, and 
{\color{brown}$\bullet$}:~$h		=20$~nm.}
\end{figure}

In order to quantify the magnitude of the platinum excess observed close to the grain-boundary groove in Fig.~\ref{slow_diff}, we measure the maximum Pt content in the system as a function of time $t$, $\cHat$. Figure~\ref{t-cMax-slow3} shows $\cHat$ as a function of time for several initial thicknesses. For thick films ($h=20$~nm), there are two regimes: $\cHat$ initially increases and then decreases for $t\ge12$~s~\footnote{For $h=20$~nm, the system size is $512 \times 192$ grid points.}. In the thinner films, $\cHat$ exhibits the same early time behavior as in the thick films, but at later times the maximum Pt concentration grows rapidly.  This film thickness effect only occurs when the Pt mobility is sufficiently low.

Since $\cHat$ exhibits the same maximum value at the same time in the thicker films (at $t \approx 12$~s), this maximum cannot be attributed to the finite thickness of the films. At early times, grooving is limited by the slow Pt diffusion; the groove can grow only as fast as it can `push' the Pt atoms out of the way. This results in a Pt build-up close to the groove. As grooving proceeds, the driving force for grooving (i.e.\ the curvature of the interface) decreases, so that at later time Pt diffusion is able to smooth the Pt concentration profile faster than groove advancement can build it.  This is what leads to the decrease of $\cHat$ around $t=12$~s in Fig.~\ref{t-cMax-slow3}.

Figure~\ref{t-cMax-slow3} shows that in thin films, the behavior of $\cHat$ is much more complex than in the thicker films.  Following the same rise in $\cHat$ as for the thick films at early times, in the thinner films, $\cHat$ decreases, then grows again, before finally decreasing.  This introduces a second maximum in $\cHat$, $\cHat\mTwo$, at later times. The first maximum in $\cHat$, $\cHat\mOne$, occurs in all thick films at the same time. The second maximum, on the other hand, depends on $h$ (for sufficiently thin films, it may occur so early that no $\cHat\mOne$ is observed).

\begin{figure}
\centering
\setlength{\unitlength}{1cm}
\begin{picture}(8.5,3)(.1,0)
\shortstack[c]{
\subfigure{
	\label{slow3(a)}
	\includegraphics[width=7.7cm]{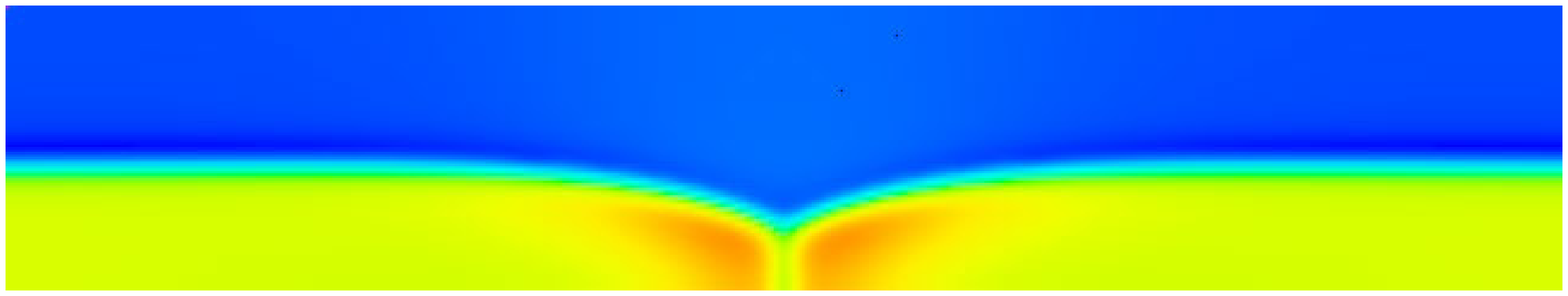}
	\put(-0.6,1.){\color{white}\bf{(a)}}
}\vspace{-3mm}\\
\subfigure{
	\label{slow3(b)}
	\includegraphics[width=7.7cm]{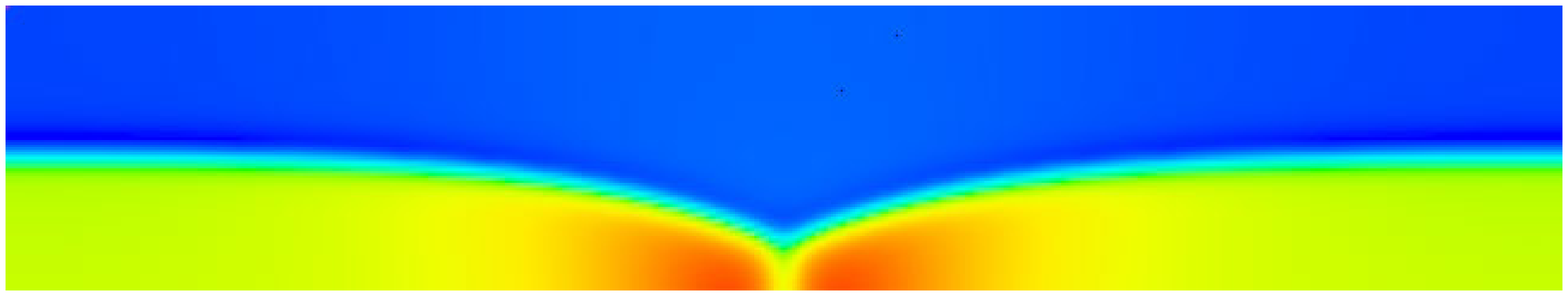}
	\put(-0.6,1.){\color{white}\bf{(b)}}
}}\;\includegraphics[height=3cm]{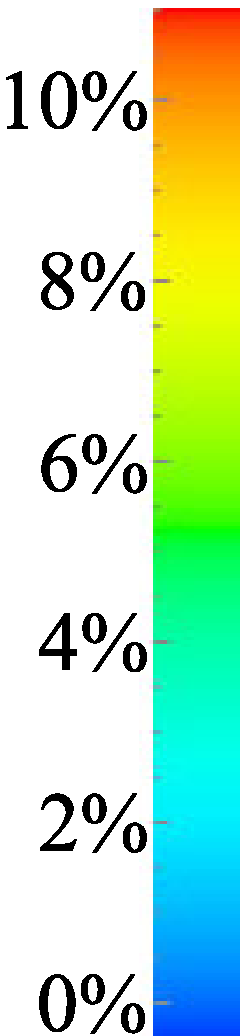}
\end{picture}
\caption{\label{slow3}(color online) The platinum composition field, $c_\text{Pt}$, for $h=10$~nm and $r = 10^{-3}$ at (a) $t=30$~s and (b) $t=1.7$~min.}
\end{figure}

In order to understand how finite thickness can affect $\cHat$, we examine the Pt composition field at different times. Along with Fig.~\ref{slow_diff3}, Fig.~\ref{slow3} shows $c_\text{Pt}$ for $h=10$~nm and $r = 10^{-3}$. Figures~\ref{slow_diff3} and~\ref{slow3(b)} correspond to the times at which the two $\cHat$ maxima in Fig.~\ref{t-cMax-slow3} occur, while Fig.~\ref{slow3(a)} corresponds to the intervening minimum. In Fig.~\ref{slow3(a)}, the region of large Pt concentration reaches the bottom of the film and in Fig.~\ref{slow3(b)}, the Pt composition is nearly uniform through the film thickness (but still depends on the distance from the grain-boundary, along~x). At this point, the regions of large Pt concentration near the groove can no longer be relaxed by diffusing in the film thickness direction (y-direction), but now must diffuse away through the film in the x-direction.  This change in transport path is very important.

\begin{figure}
\centering
\setlength{\unitlength}{1cm}
\begin{picture}(8.5,4.5)(.1,0)
\shortstack[c]{
\subfigure{
	\label{gradients(a)}
	\includegraphics[width=7.7cm]{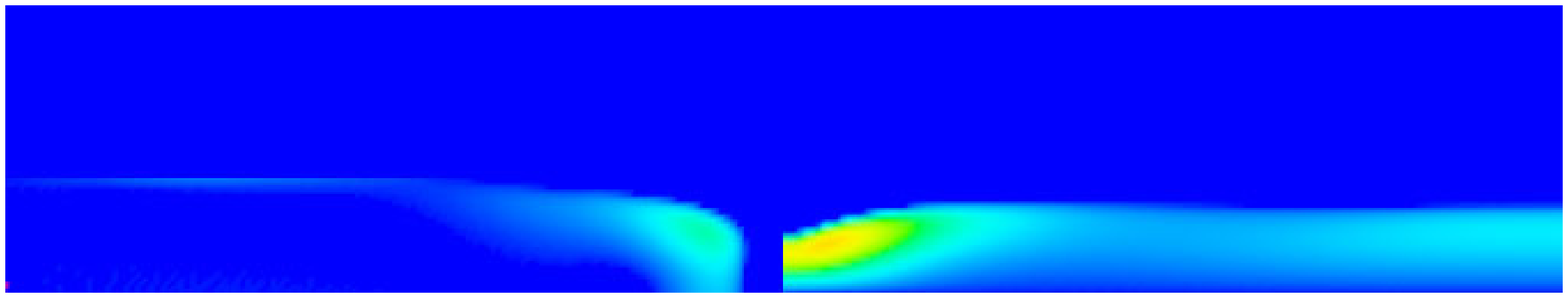}
	\put(-0.6,1.){\color{white}\bf{(a)}}
}\vspace{-3mm}\\
\subfigure{
	\label{gradients(b)}
	\includegraphics[width=7.7cm]{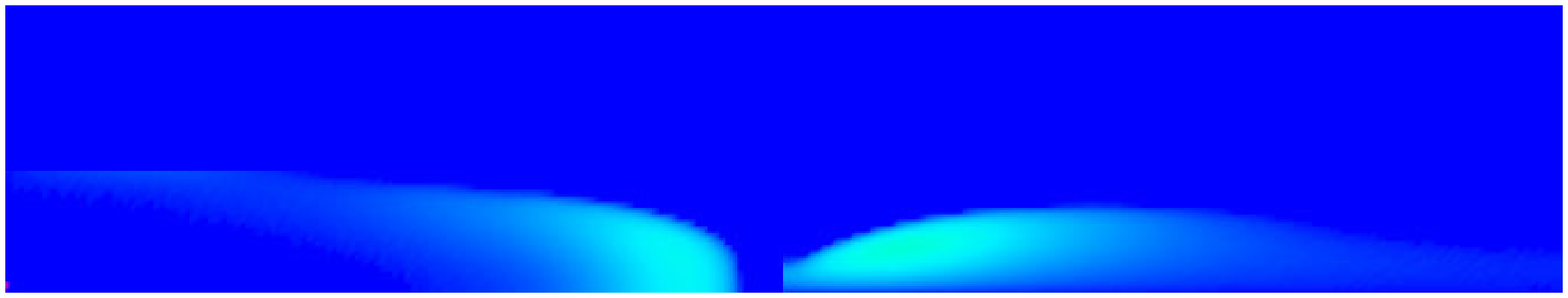}
	\put(-0.6,1.){\color{white}\bf{(b)}}
}\vspace{-3mm}\\
\subfigure{
	\label{gradients(c)}
	\includegraphics[width=7.7cm]{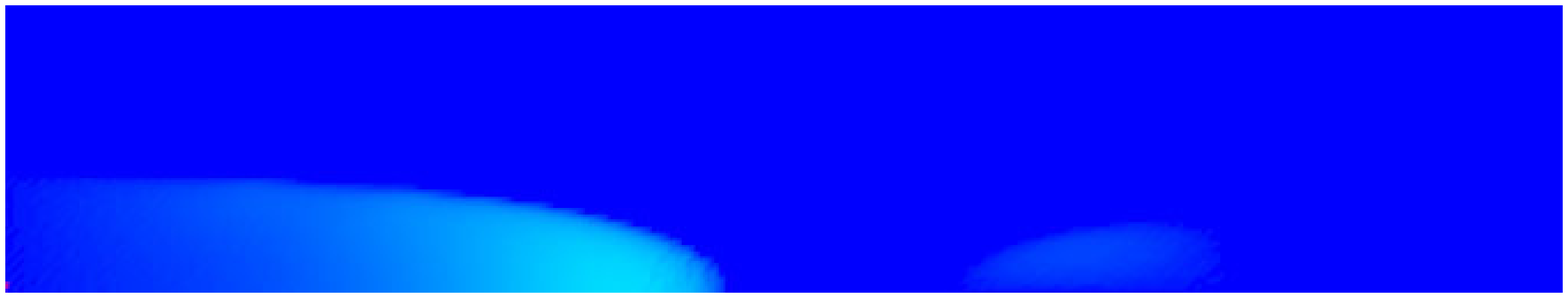}
	\put(-0.6,1.){\color{white}\bf{(c)}}
}}\;\includegraphics[height=3cm]{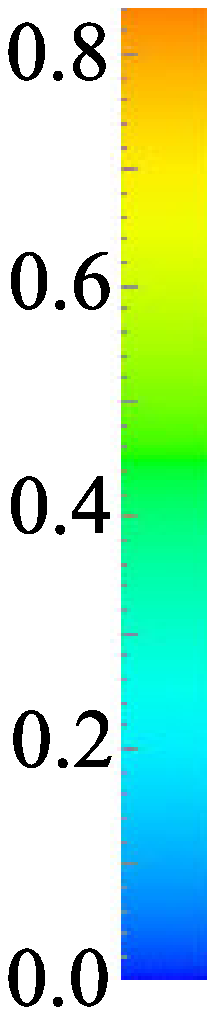}
\end{picture}
\caption{\label{gradients}(color online) Left $c'\x$ and right $c'\y$ (in \%/nm) for $h=10$~nm and $r = 10^{-3}$ at (a) $t=9$~s, (b) $t=30$~s, and (c) $t=1.7$~min. The contributions to $c'\x$ and $c'\y$ from the interface and grain-boundary have been removed for clarity.}
\end{figure}

In order to quantify the driving forces for Pt diffusion along x and y, we measure the Pt concentration gradients in the x- and y-directions, i.e.\ $c'\x = \partial c_\text{Pt}/\partial x$ and $c'\y= \partial c_\text{Pt}/\partial y$.
Figure~\ref{gradients} shows $c'\x$ and $c'\y$ at the same times as Figs.~\ref{slow_diff3} and~\ref{slow3}. It confirms that the driving force for diffusion along y drops between the two maxima of $c_\text{Pt}$, whereas the magnitude of the driving force for diffusion along x varies less over time.
While Pt transport is dominated by diffusion parallel to the grain-boundary at early time for all film thickness, in the thin films, diffusion in the y-direction is very limited (by the film thickness) at intermediate and late times.  This transition in the direction of Pt diffusion corresponds to the second rapid rise in $\cHat$ in Fig.~\ref{slow3(a)} for the thinner films. At very late times, $\cHat$  decreases as the grooving rate slows and Pt has sufficient time to diffuse away.

The times at which the exponent describing the time dependence of the groove depth, $\alpha$, changes at small~$r$ (\emph{cf.}\ Table~\ref{table-alpha}) correlate with the times at which the second maximum in $\cHat$ ($\cHat\mTwo$) occurs, as shown in Table~\ref{table-cHat}.
There is also agreement for larger values of $r$; the value of $\alpha$ is constant and there is no clear maximum of $\cHat$. Therefore, the change in the grooving exponent $\alpha$ is caused by the change in the dominant Pt transport mode.

\begin{table}
\begin{tabular}{l1lr1lr1l}
\hline
\mc{\rb{$r$}}		&\mcThree{first maximum}				&\mcFive{second maximum}\\
\mc{}				&\mcTwo{$t\mOne$}& \mc{$\cHat\mOne$}& \mcTwo{$t\mTwo$}& \mc{$\cHat\mTwo$} & \mcTwo{$d$} \\
\hline
$1\!\times\!10^0$\rule[0ex]{0ex}{2.6ex}&  2.4&ms &  8.2\% & 72&ms& 8.1\%	&  1.1	& nm\\
$1\!\times\!10^{-1}$	& 60	&ms &  8.5\% &  1.2	& s &  8.3\%	& 2.6	& nm\\
$3\!\times\!10^{-2}$	&  0.2	& s &  8.8\% &  6.0	& s &  8.8\%	& 4.1	& nm	\\
$1\!\times\!10^{-2}$	&  1.2	& s	&  9.2\% & 23	& s	&  9.6\%	& 4.8	& nm	\\
$3\!\times\!10^{-3}$	&  3.0 	& s	&  9.7\% & 42	& s	& 10.3\%	& 4.9	& nm	\\
$1\!\times\!10^{-3}$	&  9.6	& s	& 10.0\% & 1.7	& min& 10.7\%	& 5.1	& nm	\\
\hline
\end{tabular}
\caption{\label{table-cHat} The times, concentrations, and groove depth at the two maxima in $\cHat$ \emph{vs.}\ time, for $h=10$~nm.}
\end{table}

The present simulations demonstrate that introduction of a slow-diffusing species can greatly slow grain-boundary grooving and agglomeration.  The grooving leads to pile-up of the slow-diffusing species and the grooving rate depends on how quickly this excess can be diffusionally relaxed.  In the case of grain-boundary grooving in thin films, the film thickness is crucial in determining the dominant transport mode for the slow-diffusing species.  When the film is thick compared with the groove depth and the slow-diffusing species pile-up, transport is dominated by diffusion parallel to the grain-boundary.  On the other hand, when the film is thin, the transport must occur predominantly along the plane of the film.  In many cases, there is a transition in the dominant transport mode and this gives rise to a change in the exponent of the grooving power law exponent. The time at which this transition occurs depends on the film thickness and the mobility of the slow species.

We gratefully acknowledge S.~Hu and L.-Q.~Chen for their assistance with developing our phase-field simulation code. This project was funded by the A-Star Visiting Investigator Program.

\bibliography{silicides}
\bibliographystyle{apsrev}

\end{document}